\documentstyle[aps
,preprint%
]{revtex}
\newcommand{\be}{\begin{equation}}
\newcommand{\ee}{\end{equation}}
\newcommand{\bea}{\begin{eqnarray}}
\newcommand{\eea}{\end{eqnarray}}
\newcommand{\nn}{\nonumber \\}
\newcommand{\Bx}{\mbox{\boldmath$x$}}
\newcommand{\Br}{\mbox{\boldmath$r$}}
\newcommand{\Bv}{\mbox{\boldmath$v$}}
\newcommand{\BV}{\mbox{\boldmath$V$}}
\newcommand{\Bn}{\mbox{\boldmath$n$}}
\newcommand{\BN}{\mbox{\boldmath$N$}}
\newcommand{\BA}{\mbox{\boldmath$A$}}
\newcommand{\Neq}{\mbox{$\,/\!\!\!\!\!=\,$}}
\begin{document}
\draft
\preprint{AJC-HEP-26}
\date{\today
}
\title{
Many-body systems in Einstein-Maxwell-Dilaton theory}
\author{Kiyoshi~Shiraishi%
\footnote
{e-mail: {\tt g00345@sinet.ad.jp, shiraish@air.akita-u.ac.jp}
} }
\address{Akita Junior College\\
Shimokitade-sakura, Akita-shi, Akita 010, Japan
}
\maketitle
 \begin{abstract}

We study the interaction of maximally-charged dilatonic
 black holes at low velocity.
We compute the metric on moduli space for three extreme black holes
under a simple constraint.
The Hamiltonian of the multi-black hole system of $O(v^2)$
is also calculated for the $a=1$ and $a=1/\sqrt{3}$ cases,
where $a$ is the dilaton coupling constant.
The behavior of the system is discussed qualitatively.

\end{abstract}
\vspace{7mm}
\pacs{PACS number(s): 04.20.Cv,
04.20.Me, 
97.60.Lf}

It has been considered
for recent years that solitonic objects
 play an important role in theoretical physics.
Static multi-soliton solutions have been found in many cases and
their physical implication has been studied by many authors.
There is no net static force among the solitonic objects described
by such a multi-soliton solution.

The slow motion of such solitons can be described as
 a sequence of static solutions for every moment.
 The evolution of the solution is then approximated
 in terms of geodesics on the space of parameters (moduli space)
 for static solutions.%
\footnote{In this analysis, radiation reaction is neglected.}
Therefore the calculation of the metric on moduli space, which
 defines geodesic motion
on the moduli space, is the most effective way to
investigate interaction of slowly moving solitons
in classical field theory~\cite{Manton}.

The present author studied the interaction Hamiltonian of maximally
charged dilatonic black holes and obtained the moduli space metric
for two extreme black holes explicitly~\cite{Shi2,Shi3}.

In the present paper, we will first
examine a three-body problem of the extreme
black holes at low velocity in a simple configuration.
The metric on moduli space for three extreme black holes will be
calculated in this case.
Next we will consider arbitrarily many body problems for $a=1$
and $a=1/\sqrt{3}$,
where $a$ is the dilaton coupling constant.
We will compute the Hamiltonian of the multi-black hole system
up to $O(v^2)$ for the two cases.


We suppose the action for the Einstein-Maxwell-dilaton theory is
given by
\be
S=\int d^{4}x \frac{\sqrt{-g}}{16\pi} \left[ R -
2 (\nabla \phi)^{2} - e^{-2a\phi} F^{2} \right] +
\mbox{(surface terms)}\; ,
\ee
where we set the Newton's constant $G=1$.
$F_{\mu\nu}$ is the field strength of the electromagnetic field.
The dilaton coupling constant $a$
 can be assumed to be a positive value.

The static multi-centered solution for the metric, vector potential one-
form,
 and dilaton field takes the form~\cite{GHS,Shi1}
\be
ds^{2}= - \frac{dt^{2}}{V^{2/(1+a^{2})}} + V^{2/(1+a^{2})} d\Bx^{2},\;
A =  - \frac{1}{\sqrt{1+a^{2}}\ V} dt,\;
e^{-2\phi}=V^{2a^{2}/(1+a^{2})}\; ,
\label{eq:met}
\ee
where
\be
V=1+\sum_{\alpha} \frac{(1+a^{2}) m_{\alpha}}{|\Bx -\Bx_{\alpha}|}\; .
\ee
The metric corresponds to the situation that the $\alpha$-th
nonrotating, extreme black hole with mass
 $m_{\alpha}$
is located at $\Bx =\Bx_{\alpha}$.%
\footnote{Strictly speaking, for $a \Neq 0$, there are
naked singularities in the solution.
Nevertheless, we use the term ``black hole'' because the extreme case
may still have generic properties of black holes in terms of
such classical dynamics.}
 The electric repulsion is balanced by the
gravitational force and the attractive dilatonic force.
Therefore this solution can be regarded as an extension of
the Majumdar-Papapetrou solution~\cite{MP}.

We apply the method of Ferrell and Eardley~\cite{FerEar} to calculate
the velocity-dependent
 interaction energy of the slow motion of the extreme black holes
by making use of the static solution~(\ref{eq:met}).

We take the perturbed metric and potential of the form
\be
ds^{2}= - \frac{dt^{2}}{V^{2/(1+a^{2})}} + 2\BN\cdot d\Bx dt +
V^{2/(1+a^{2})} d\Bx^{2},\;
A=  - \frac{1}{\sqrt{1+a^{2}}\ V}dt + \BA\cdot d\Bx\; .
\ee
We solve linearized equations for $\BN$ and $\BA$
coupled to the slowly moving
 extreme black holes as sources of gravitation and electromagnetism.
 The solution is substituted to
the action to get the interaction energy of extreme black holes.
The equations in detail for its derivation
can be referred to the previous work~\cite{Shi2,Shi3}.

Consequently, the effective Hamiltonian of the order of $v^2$ for the
 maximally-charged-dilatonic-black-hole system is obtained
 as~\cite{Shi2,Shi3}
\be
H=\sum_{\alpha}\frac{1}{2}m_{\alpha}\Bv_{\alpha}^{2} +
\frac{3-a^{2}}{8\pi} \int d^{3}\Bx\: V(\Bx)^{2(1-a^{2})/(1+a^{2})}
\sum_{\alpha\beta}
\frac{m_{\alpha}m_{\beta} (\Bn_{\alpha}\cdot\Bn_{\beta})
 |\Bv_{\alpha}-\Bv_{\beta}|^{2}}
{2|\Br_{\alpha}|^{2} |\Br_{\beta}|^{2}}\; ,
\label{eq:H}
\ee
where $\Br_{\alpha}\equiv\Bx-\Bx_{\alpha}$ and
$\Bn_{\alpha}\equiv\Br_{\alpha}/|\Br_{\alpha}|$.

In the previous papers~\cite{Shi2,Shi3} we investigated
two-body system and computed the metric on the moduli space
from the expression~(\ref{eq:H}).
In the present paper, we first study a constrained three-body system
and its moduli space.

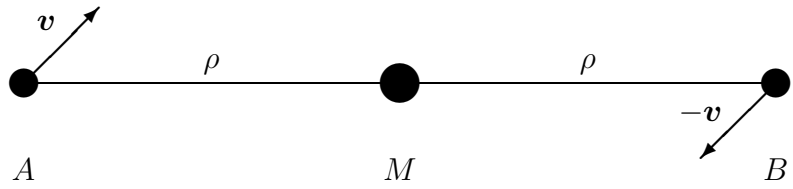
\begin{figure}
\unitlength 1mm
\begin{picture}(150,80)(-75,-40)
\put(50,10){\circle*{4}}
\put(0,10){\circle*{8}}
\put(-50,10){\circle*{4}}
\put(50,0){\makebox(0,0)[ct]{$B$}}
\put(0,0){\makebox(0,0)[ct]{$M$}}
\put(-50,0){\makebox(0,0)[ct]{$A$}}
\put(-25,12){\makebox(0,0)[cb]{$\rho$}}
\put(25,12){\makebox(0,0)[cb]{$\rho$}}
\put(-50,10){\line(1,0){100}}
\put(40,5){\makebox(0,0)[cb]{$-\Bv$}}
\put(-47,17){\makebox(0,0)[cb]{$\Bv$}}
\thicklines
\put(50,10){\vector(-1,-1){10}}
\put(-50,10){\vector(1,1){10}}
\end{picture}
\caption{The configuration of the three black holes.}
\label{fig1}
\end{figure}

The simplest configuration we consider
is depicted as in FIG.~\ref{fig1}.
Three points $O$, $A$, and $B$
lie on the same straight line. Moreover, the distance $OA$ is set to be
 equal to that of $OB$, which is denoted as $\rho$.
We locate the center of extreme black holes
of the same mass $m$ at the point $A$ and $B$,
 and one of the mass $M$ at
 the point $O$. Further we assume that
the velocity of the black hole at $A$
is $\Bv$ and that of the black hole at $B$ is $-\Bv$.
Putting these assumptions, we observe that the center of mass $O$ is
 fixed and the linear configuration is not broken due to the symmetry.
In addition, the motion of the black holes is assumed to be constrained
in a (scattering) plane.
Thus the moduli space of this configuration
is reduced to be a two dimensional space,
parameterized by the distance $\rho$ and
the azimuthal angle $\varphi$.

For this configuration, the Hamiltonian will be written in the form
\be
H=m\, \gamma(\rho)\, \Bv^{2}\; .
\label{eq:H3}
\ee
Then the metric on moduli space can be read from (\ref{eq:H3}) as
\be
ds_{MS}^2=\gamma(\rho)\left(d\rho^{2}+\rho^{2}d\varphi^{2}\right)\; .
\ee

Explicit calculation of $H$ (eq.~(\ref{eq:H})) yields the moduli space
metric. We calculate the metric for a few cases for special values
of the dilaton coupling $a$.

For the case of the system of extreme Reissner-Nordstr\"om
black holes
with $a=0$, $\gamma$ is obtained from the explicit calculation of
 $H$ (eq.~(\ref{eq:H})) for this symmetric configuration:
\be
\gamma(\rho)=1+3\, \frac{m+M}{\rho}+3\, \frac{(m+M)^{2}}{\rho^{2}}+
\frac{\frac{m^{3}}{2}+m^{2}M+M^{3}}{\rho^{3}} \qquad (a=0)\; .
\ee

For the case with $a=1/\sqrt{3}$, which corresponds
to the reduction of five-dimensional Einstein-Maxwell or
supergravity theory to four dimensions, we find
\be
\gamma(\rho)=\left(1+\frac{4\, (m+M)}{3\, \rho}\right)^{2} \qquad
(a=1/\sqrt{3})\;.
\ee

For the stringy case $a=1$, it turns out that
\be
\gamma(\rho)=1+2\, \frac{m+M}{\rho} \qquad (a=1)\;.
\ee

For $a=\sqrt{3}$, which corresponds
to the reduction of five-dimensional Einstein gravity
to four dimensions, the leading order metric on moduli space is flat,
i.~e. $\gamma = 1$.
This implies that to leading order the dynamical force vanishes
identically when $a=\sqrt{3}$
and the scattering is trivial in this case.

In comparison with the previous work~\cite{Shi3},
the global nature of moduli space is found to be the same as that
for the two-body system in each case,
since the coefficient of each order of
$\rho$ is {\em positive}.
The schematic view of the moduli space is shown in FIG.~\ref{fig2}.

For general configuration for three-body system is difficult to treat
analytically, for general dilaton coupling.
There are however two cases, for $a=1$ and for $a=1/\sqrt{3}$,
in which the integration in (\ref{eq:H}) can be carried out.

Since there is only two-body (velocity-dependent) force
in the multi-black hole system for
$a=1$,
the general expression for the $O(v^2)$ Hamiltonian of the system
of an arbitrary configuration of black holes
can be easily written down:
\bea
H&=&\sum_{\alpha}\frac{1}{2}m_{\alpha}\Bv_{\alpha}^{2} +
\sum_{\alpha\beta}
\frac{m_{\alpha}m_{\beta} |\Bv_{\alpha}-\Bv_{\beta}|^{2}}
{2\rho_{\alpha\beta}}\nn
&=&\frac{1}{2}M\BV^{2} +
\sum_{\alpha\beta}
\frac{m_{\alpha}m_{\beta} |\Bv_{\alpha}-\Bv_{\beta}|^{2}}{4 M}
\left(1+\frac{2 M}{\rho_{\alpha\beta}}\right)\;,
\label{eq:ie1}
\eea
where $\rho_{\alpha\beta}\equiv |\Bx_{\alpha}-\Bx_{\beta}|$,
$M=\sum_{\alpha} m_{\alpha}$,
 and $\BV$ is the velocity of
the center of mass; $\BV\equiv
\frac{\sum_{\alpha} m_{\alpha} \Bv_{\alpha}}{M}$.

Similarly, for the $a=1/\sqrt{3}$ case, in which there are
two- and three-body
 forces among the extreme black holes,
 the energy of the system can be expressed as
\bea
H&=&\sum_{\alpha}\frac{1}{2}m_{\alpha}\Bv_{\alpha}^{2} +
\frac{2}{3}
\sum_{\alpha\beta}
\frac{m_{\alpha}m_{\beta} |\Bv_{\alpha}-\Bv_{\beta}|^{2}}
{\rho_{\alpha\beta}}\nn
{}&+&\frac{4}{9}
\sum_{\alpha\beta\gamma}
m_{\alpha}m_{\beta}m_{\gamma} |\Bv_{\alpha}-\Bv_{\beta}|^{2}
\left(
\frac{1}{\rho_{\alpha\beta}\rho_{\alpha\gamma}}+
\frac{1}{\rho_{\alpha\beta}\rho_{\beta\gamma}}-
\frac{1}{\rho_{\alpha\gamma}\rho_{\beta\gamma}}
\right)\nn
&=&\frac{1}{2}M\BV^{2}\nn
{}&+&\sum_{\alpha\beta\gamma}
\frac{m_{\alpha}m_{\beta}m_{\gamma} %
|\Bv_{\alpha}-\Bv_{\beta}|^{2}}{4 M^{2}}
\left[1+
\frac{8}{3} \frac{M}{\rho_{\alpha\beta}} +
\frac{16 M^{2}}{9} \left(
\frac{1}{\rho_{\alpha\beta}\rho_{\alpha\gamma}}+
\frac{1}{\rho_{\alpha\beta}\rho_{\beta\gamma}}-
\frac{1}{\rho_{\alpha\gamma}\rho_{\beta\gamma}}
\right)
\right]\;.
\label{eq:ie1/3}
\eea
In the two cases above, no regularization on the integration is needed.

It is difficult, however, to analyze a general many-body problem of
extreme black holes even if one uses the Hamiltonian
 (eq.~(\ref{eq:ie1},\ref{eq:ie1/3})).
Thus we qualitatively examine the slow motion of the black holes here.

Let us consider
that two black holes labelled by $A$ and $B$ come close to each other.

For $a=1$,
the divergent term in the Hamiltonian (\ref{eq:ie1})
in the limit $\rho_{AB}\rightarrow 0$ behaves
 as $\approx 1/\rho_{AB}$.
Accordingly, each pair of black holes will interact like the
Rutherford scattering~\cite{Shi3}.
Hence the coalescence of black holes is
unlikely to occur in general conditions in the stringy case $a=1$.

For $a=1/\sqrt{3}$,
the divergent term in the Hamiltonian (\ref{eq:ie1/3})
 behaves as $\approx 1/\rho_{AB}^{2}$
in the limit $\rho_{AB}\rightarrow 0$.
In this case, the coalescence of black holes occurs
if black holes come sufficiently close.
Actually, this case is considered to be critical:
For $a>1/\sqrt{3}$ extreme black holes never coalesce~\cite{Shi3}.

In this paper we have studied the many-body system of extreme
black holes in Einstein-Maxwell-dilaton theory,
in the low-velocity limit.
As an example, a simple three-body case has been studied
 and the metric on its moduli space has been obtained.
We have found that the moduli space is very similar to that for
two-body system.
For the cases with $a=1$ and $a=1/\sqrt{3}$, the Hamiltonian
for many-body system of extreme black holes has been
explicitly obtained.
The qualitative nature of the systems has been discussed,
but the detailed analysis has been left for the future work,
in which one may demand numerical simulations to reveal
the property of the system, such as
appearance of chaotic behavior~\cite{chaos}.


\newpage

\newpage

\begin{figure}
\unitlength 1mm
\begin{picture}(150,200)(-75,-100)
\end{picture}
\caption{Schematic view of the surface of the moduli space.}
\label{fig2}
\end{figure}
\begin{center}
(a) for $a=0$, (b) for $a=1/\sqrt{3}$, and (c) for $a=1$.
\end{center}


\begin{references}

\bibitem{Manton}
N.~Manton, Phys.\ Lett.\ {\bf B110}, 54 (1982); {\it ibid.} {\bf B154},
397 (1985).\\
R.~S.~Ward, Phys.\ Lett.\ {\bf B158}, 424 (1985).\\
M.~F.~Atiyah and N.~J.~Hitchin,
{\em The Geometry and Dynamics of Magnetic Monopoles}
(Princeton University Press, Princeton, 1988).

\bibitem{Shi2} K.~Shiraishi, Nucl.\ Phys.\ {\bf B402}, 399 (1993).

\bibitem{Shi3} K.~Shiraishi, Int.\ J.\ Mod.\ Phys.\
{\bf D2}, 59 (1993).

\bibitem{GHS}
D.~Garfinkle, G.~Horowitz and A.~Strominger, Phys.\ Rev.\
{\bf D43}, 3140 (1991);
 {\it ibid.} {\bf D45}, 3888 (E) (1992).

\bibitem{Shi1} K.~Shiraishi, J.\ Math.\ Phys.\ {\bf 34}, 1480 (1993).

\bibitem{MP}
A.~Papapetrou, Proc.\ R.\ Irish Acad.\ {\bf A51}, 191 (1947).\\
S.~D.~Majumdar, Phys.\ Rev.\ {\bf 72}, 930 (1947).\\
R.~C.~Myers, Phys.\ Rev.\ {\bf D35}, 455 (1987).

\bibitem{FerEar}
R.~C.~Ferrell and D.~M.~Eardley, Phys.\ Rev.\ Lett.\ {\bf 59},
1617 (1987).\\
J.~Traschen and R.~Ferrell, Phys.\ Rev.\ {\bf D45}, 2628 (1992).

\bibitem{chaos}
C.~P.~Dettmann, N.~E.~Frankel and N.~J.~Cornish,
Phys.\ Rev.\ {\bf D50}, R628 (1994); preprint UTPT-94-22,
UM-P-94/87, gr-qc/9502014.\\
U.~Yurtsever, preprint gr-qc/9412031, and references therein.
\\
\\
\\
\end{references}
\end{document}